# Red alert: Millions of "homeless" publications in Scopus should be resettled

Weishu Liu    wsliu08@163.com
https://orcid.org/0000-0001-8780-6709
School of Management, Zhejiang University of Finance and Economics, Hangzhou 310018, Zhejiang, China

Haifeng Wang    corresponding author    wanghf@shisu.edu.cn
https://orcid.org/0000-0003-4967-7971
School of Business and Management, Shanghai International Studies University, Shanghai 200083, China

**Abstract** Scopus is increasingly regarded as a high-quality and reliable data source for research and evaluation of scientific and scholarly activity. However, a puzzling phenomenon has been discovered occasionally: millions of records with author affiliation information collected in Scopus are oddly labeled as "country-undefined" by Scopus which is rarely to be detected in its counterpart Web of Science. This huge number of "homeless" records in Scopus is unacceptable for a widely used high-quality bibliographic database. By using data from the past 124 years, this brief communication tries to probe these affiliated but country-undefined records in Scopus. Our analysis identifies four primary causes for these "homeless" records: incomplete author affiliation addresses, Scopus' inability to recognize different variants of country/territory names, misspelled country/territory names in author affiliation addresses, and Scopus' insufficiency in correctly split and identify the clean affiliation addresses. To address this pressing issue, we put forward several recommendations to relevant stakeholders, with the aim of resettling millions of "homeless" records in Scopus and reducing its potential impact on Scopus-based literature retrieval, analysis, and evaluation.

## 1. Introduction

As multidisciplinary bibliographic databases, both the established Web of Science and the rising power Scopus self-label as high-quality and reliable data sources for research and evaluation of scientific and scholarly activity (Baas et al., 2020; Birkle et al., 2020). These two databases are also widely adopted by researchers and evaluators as reliable data sources for literature retrieval, analysis, and evaluation (Zhu & Liu, 2020)[1]. However, we recently encountered a puzzling phenomenon when utilizing Scopus for literature retrieval and analysis. Specifically, a significant number of records affiliated with distinguished universities, such as the Massachusetts Institute of Technology and the University of Cambridge, are labeled as "country-undefined" in Scopus. In contrast, a comparative analysis of the Web of Science Core Collection, conducted through multiple rounds of testing, does not reveal a similar occurrence. These affiliated yet country-undefined records may be omitted when searching the country/territory names in affiliation or affiliation country fields in Scopus. Furthermore, if these records are retrieved using alternative search strategies, they may be excluded from country-level analyses. These two scenarios pose nonnegligible threats to the reliability of many Scopus-based literature retrieval, analyses, and evaluations.

---

[1] For example, the famous biennial report Science and Engineering Indicators uses data from Scopus to identify Science & Engineering articles by selected region, country, or economy. More information can be accessed at https://ncses.nsf.gov/pubs/nsb20243.

Previous studies have documented that many country-undefined records exist in Scopus and Web of Science Core Collection (Jacsó, 2009; Savchenko & Kosyakov, 2022). Studies also propose methods, including using publication languages and the affiliation history of the authors, to predict the affiliation country/territory information in these records (Mryglod & Nazarovets, 2023; Savchenko & Kosyakov, 2022). However, it is generally accepted that the null affiliation country/territory field in these established bibliographic databases is indicative of the absence of author affiliation address information which has been probed by Liu et al. (2018) focusing on Web of Science Core Collection. For instance, records published anonymously typically lack author affiliation address information (Li & Zhang, 2024; Shamsi et al., 2022). However, according to the observed phenomenon in Scopus, it is imprecise to equate country-undefined records to records without affiliation address information in Scopus. In contrast to records without country/territory information in Web of Science, many of these country-undefined records in Scopus have affiliation address information collected in the database. For example, our data reveals that over one million records with affiliation address information collected in Scopus are oddly labeled as "country-undefined" by Scopus.

Millions of affiliated but country-undefined records are unacceptable for a widely used high-quality bibliographic database. In this brief communication, we try to probe these "homeless" publications in Scopus and identify the potential causes for the imprecise labeling. Similar checks are conducted in the Web of Science Core Collection for comparison. Finally, we endeavor to offer stakeholders, including bibliographic database providers, academic publishers, and database users, potential solutions to address this pressing issue.

## 2. Data and methods

A typical author affiliation address is formatted as "School name, University name, City, State, Country". A string comprises alphanumeric characters ranging from A to Z and 0 to 9 with the aid of the wildcard "*" as follows is employed to ascertain whether the value of a designated field in Scopus and Web of Science Core Collection is null or not (Liu et al., 2018). Records with non-null values identified in the author address field in Web of Science Core Collection (field tag: AD) and the affiliation field in Scopus (field tag: AFFIL) are called "affiliated records". A combination of the affiliation field and affiliation country field (field tag: AFFILCOUNTRY) with the operator "AND NOT" can be used to retrieve affiliated but country-undefined records in Scopus[2]. However, it is important to note that the search field AFFILCOUNTRY provided by Scopus does not apply to the corresponding address field (Huang et al., 2023). To identify country-undefined records in Scopus, we employ the affiliation country/territory filter tool (i.e., LIMIT-TO ( AFFILCOUNTRY , "Undefined" )). For Web of Science Core Collection, a combination of the address field and country/region field (field tag: CU) with the operator "NOT" is used to retrieve affiliated but country-undefined records in Web of Science Core Collection. The timespan is set as 1900-2023 (i.e., Scopus: PUBYEAR > 1899 AND PUBYEAR < 2024; Web of Science Core Collection: FPY=1900-2023)[3]. The initial data were accessed on November 27, 2024, and the subsequent update was executed on January 10, 2025.

---

[2] A bit different from the author address field (field tag: AD) in Web of Science Core Collection, the affiliation field (field tag: AFFIL) in Scopus is a combined field that searches in the organization, city, and country portions of the author affiliation address field.

Search query for Scopus: AFFIL (A* OR B* OR C* OR D* OR E* OR F* OR G* OR H* OR I* OR J* OR K* OR L* OR M* OR N* OR O* OR P* OR Q* OR R* OR S* OR T* OR U* OR V* OR W* OR X* OR Y* OR Z* OR 0* OR 1* OR 2* OR 3* OR 4* OR 5* OR 6* OR 7* OR 8* OR 9*) AND LIMIT-TO ( AFFILCOUNTRY , "Undefined" )

Search query for Web of Science Core Collection: AD=(A* OR B* OR C* OR D* OR E* OR F* OR G* OR H* OR I* OR J* OR K* OR L* OR M* OR N* OR O* OR P* OR Q* OR R* OR S* OR T* OR U* OR V* OR W* OR X* OR Y* OR Z* OR 0* OR 1* OR 2* OR 3* OR 4* OR 5* OR 6* OR 7* OR 8* OR 9*) NOT CU=(A* OR B* OR C* OR D* OR E* OR F* OR G* OR H* OR I* OR J* OR K* OR L* OR M* OR N* OR O* OR P* OR Q* OR R* OR S* OR T* OR U* OR V* OR W* OR X* OR Y* OR Z* OR 0* OR 1* OR 2* OR 3* OR 4* OR 5* OR 6* OR 7* OR 8* OR 9*)

## 3. Affiliated but country-undefined publications in Scopus

According to Scopus, among a total of 94.5 million records published from 1900 to 2023, 16.0% (15.1 million) of them are labeled as country-undefined records. Notably, among these country-undefined records, 12.4% (nearly 1.9 million) of them include affiliation information in Scopus. Comparatively, among 89.9 million records indexed in the Web of Science Core Collection for the same period, 16.8 % (15.1 million) of them are country-undefined records. Of the huge number of country-undefined records in the Web of Science Core Collection, only 11 of them are accompanied by affiliation address information in the database[4].

Figure 1 illustrates the dynamics of the affiliated but country-undefined records in Scopus over the past 124 years. The figure indicates a consistent presence of these records throughout the 124 years, with a notable peak around the year 1985. For the year 1985, approximately 135,000 affiliated records in Scopus were designated as country-undefined. A staged peak also happened during the past five years between 2019 and 2023, with an annual average of approximately 20,000 affiliated but country-undefined records. Since the gradual expansion of Scopus, the share of affiliated but country-undefined records among all Scopus-indexed records has also been calculated for comparison. During the period from 1900 to 1982, approximately 4% of each year's Scopus-indexed records were affiliated but country-undefined. For the peak period around the year 1985, particularly in 1985, 16.7% of the Scopus-indexed records were affiliated but country-undefined. After the peak period, a downward trend can be observed, with the affiliated but country-undefined records maintaining a share of less than 1% of the total records in recent years.

---

[3] Due to the delay in indexing, we do not include the year 2024 in our analysis.
[4] We don't have a full subscription to the Web of Science Core Collection; the subscribed and used sub-datasets of Web of Science Core Collection including SCIE and SSCI from 1900, A&HCI from 1975, CPCI-S and CPCI-SSH from 1990, BKCI-S and BKCI-SSH from 2005, and ESCI from 2015.

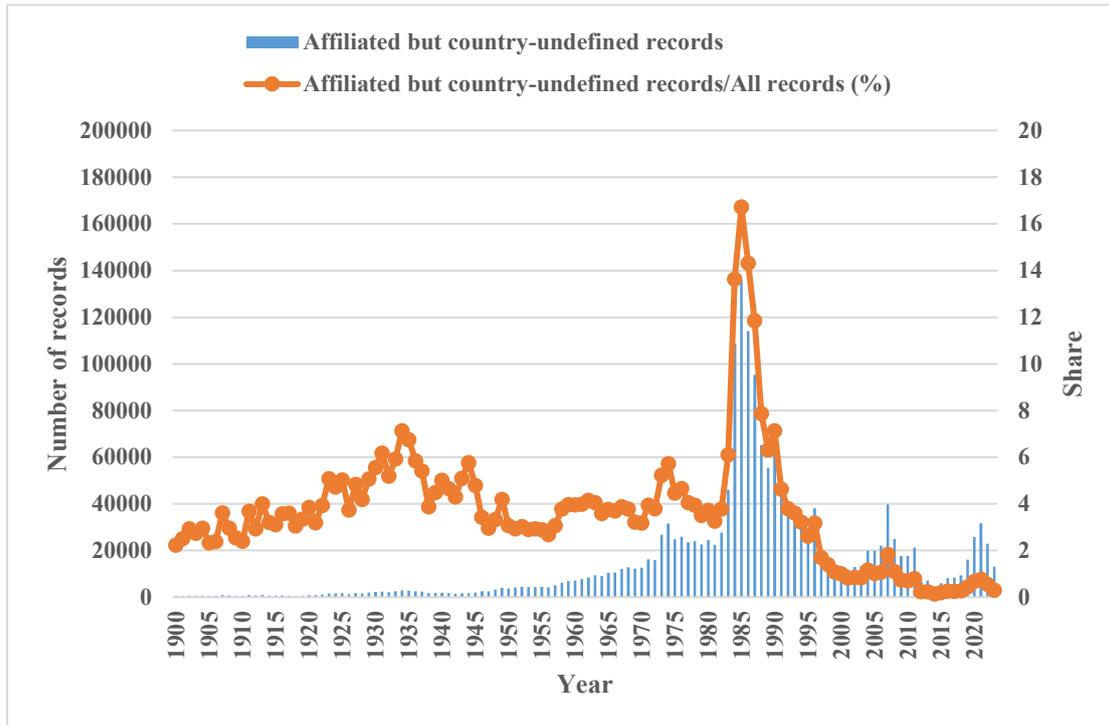

Figure 1 Dynamics of affiliated but country-undefined records (all document types)

We further choose two widely used document types, articles and reviews, for further analysis. According to Scopus, among a total of 71.3 million articles and reviews published from 1900 to 2023, 14.9% (10.6 million) of them were labeled as country-undefined. About 12.2% (1.3 million) of these country-undefined articles and reviews have affiliation information collected in Scopus. Figure 2 provides a comprehensive representation of the dynamics of affiliated yet country-undefined articles and reviews over the past 124 years, offering both absolute and relative perspectives. Similar to Figure 1 focusing on all Scopus-indexed document types, Figure 2 reveals analogous trends for Scopus-indexed articles and reviews.

The affiliated yet country-undefined records are not concentrated on a few organizations. A considerable number of affiliation organizations have published a substantial quantity of affiliated but country-undefined records according to Scopus. Table 1 presents the top 20 affiliation organizations with the largest numbers of affiliated but country-undefined records for the 124-year period, considering all document types. Of particular note are three leading organizations from the Russian Federation, namely the Russian Academy of Sciences (25,321 records), Lomonosov Moscow State University (7,310 records), and the Russian Academy of Medical Sciences (7,275 records). However, these three organizations only account for approximately 1.36%, 0.39%, and 0.39% of all affiliated but country-undefined records in Scopus. As illustrated in Table 1, a significant proportion of these affiliated yet country-undefined records are attributable to researchers based in countries such as the United States, the United Kingdom, France, and the Czech Republic[5]. Similar findings can be detected if only articles and reviews are considered.

---

[5] The country/territory information is accessed from the organization database of Scopus. Some organizations may have overseas branches. Therefore, a proportion of records affiliated with these affiliation organizations may belong to other countries.

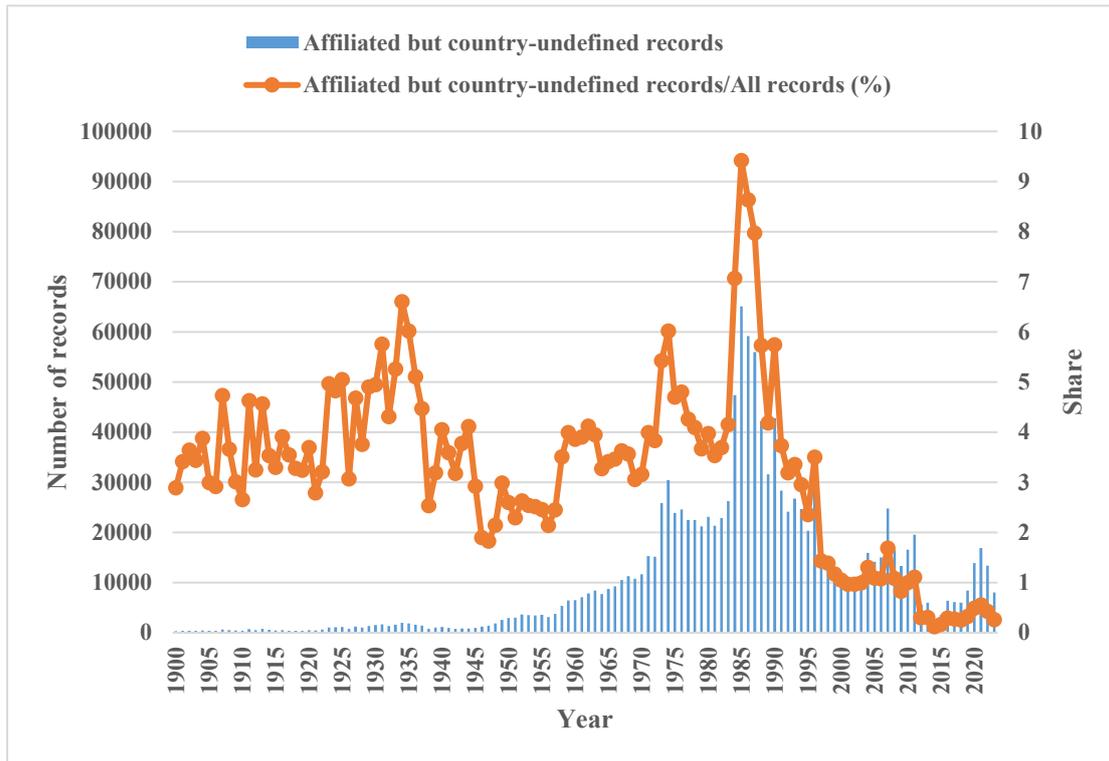

Figure 2 Dynamics of affiliated but country-undefined records (articles and reviews)

Table 1 Main organizations with affiliated but country-undefined records (all document types)

| Rank | Organization | Affiliated but country-undefined records | Share among all affiliated but country-undefined records (%) | Country/Territory |
|---|---|---|---|---|
| 1 | Russian Academy of Sciences | 25321 | 1.36 | Russian Federation |
| 2 | Lomonosov Moscow State University | 7310 | 0.39 | Russian Federation |
| 3 | Russian Academy of Medical Sciences | 7275 | 0.39 | Russian Federation |
| 4 | CNRS Centre National de la Recherche Scientifique | 6964 | 0.37 | France |
| 5 | Imperial College London | 6909 | 0.37 | United Kingdom |
| 6 | University of Oxford | 6815 | 0.37 | United Kingdom |
| 7 | Massachusetts Institute of Technology | 6759 | 0.36 | United States |
| 8 | Nokia Bell Labs | 6649 | 0.36 | United States |
| 9 | Charles University | 6457 | 0.35 | Czech Republic |
| 10 | University of Cambridge | 6063 | 0.32 | United Kingdom |
| 11 | Stanford University | 5969 | 0.32 | United States |
| 12 | University of California, Berkeley | 5845 | 0.31 | United States |
| 13 | University of Illinois Urbana-Champaign | 5660 | 0.30 | United States |
| 14 | The University of Manchester | 5622 | 0.30 | United Kingdom |
| 15 | Academy of Sciences of the Czech Republic | 5314 | 0.28 | Czech Republic |
| 16 | University College London | 5062 | 0.27 | United Kingdom |
| 17 | National Academy of Sciences of Ukraine | 5051 | 0.27 | Ukraine |
| 18 | Akademie der Wissenschaften der DDR | 5000 | 0.27 | Germany |
| 19 | University of Michigan, Ann Arbor | 4850 | 0.26 | United States |
| 20 | Cornell University | 4594 | 0.25 | United States |

## 4. Causes of affiliated but country-undefined publications in Scopus

Compared to only 11 affiliated but country-undefined publications in the Web of Science Core Collection, millions of corresponding records in Scopus shock us. A careful examination of the data reveals four salient factors that contribute to the substantial presence of affiliated but country-undefined in the Scopus database.

The primary and most evident cause is the incompleteness of the author affiliation address fields, which is characterized by the absence of the affiliation country/territory portion in Scopus. To illustrate this phenomenon, an examination of 6,964 country-undefined publications affiliated with CNRS Centre National de la Recherche Scientifique reveals that 45.9% of them lack the country name "France" in the affiliation address field. A significant contributing factor is the tendency of some authors to provide only part of the author affiliation address information in their publications (e.g., the affiliation organization and city information), often omitting the affiliation country/territory information. However, to address these kinds of publications, Clarivate's Web of Science Core Collection has employed a method of inference to supplement the affiliation country/territory information based on the existing incomplete affiliation addresses (Liu et al., 2018). This positive and responsible action explains the scarcity of affiliated but country-undefined records within the Web of Science Core Collection.

The second representative cause is Scopus' incapability in recognizing different variants of country/territory names. A significant number of affiliated yet country-undefined records contain affiliation country/territory information in the affiliation fields in Scopus. However, the country/territory names are not provided in a "standard" way that Scopus can recognize. A notable example is the fact that a significant number of records containing "USA" in their affiliation fields are also labeled as country-undefined by Scopus. For instance, employing the search query "AFFIL(USA) AND PUBYEAR > 1899 AND PUBYEAR < 2024" retrieves nearly 440 thousand records in Scopus. However, 301 thousand (68.6%) of them have been classified as country-undefined[6]. A subsequent examination of the top 20,000 records, ranked by the number of citations, reveals that the majority of these records contain the term "USA" in the country portion of the affiliation fields in Scopus. One potential explanation for this phenomenon is that Scopus fails to equate the country name "USA" with its recognized country name "United States"[7]. Figure 3 provides a visual representation of the dynamics of records with "USA" in the affiliation field and also those labeled as country-undefined[8]. A significant peak in the number of records with "USA" in the affiliation field also appeared around the year 1985. Maybe for some unknown reasons, Scopus neglected to transform the affiliation country "USA" of many records in this period to "United States". However, no solid evidence has been found yet. A similar peak also appeared around the year 1985 for country-undefined records with "USA" in the affiliation field. It offers a potential explanation for the observed peak around the year 1985 in Figure 1.

---

[6] About 134,000 records are attributed to the affiliation country "United States" by Scopus since they contain "United States" in the affiliation country or correspondence address field.
[7] Comparatively, the standard country name for the United States of America in the Web of Science Core Collection is "USA".
[8] By using the Web of Science Core Collection, Liu et al. (2018) find many records contributed by the USA for the period of 1966-1998 are without "USA" in the address field. However, the missing "USA" has been added to the country field by the database provider.

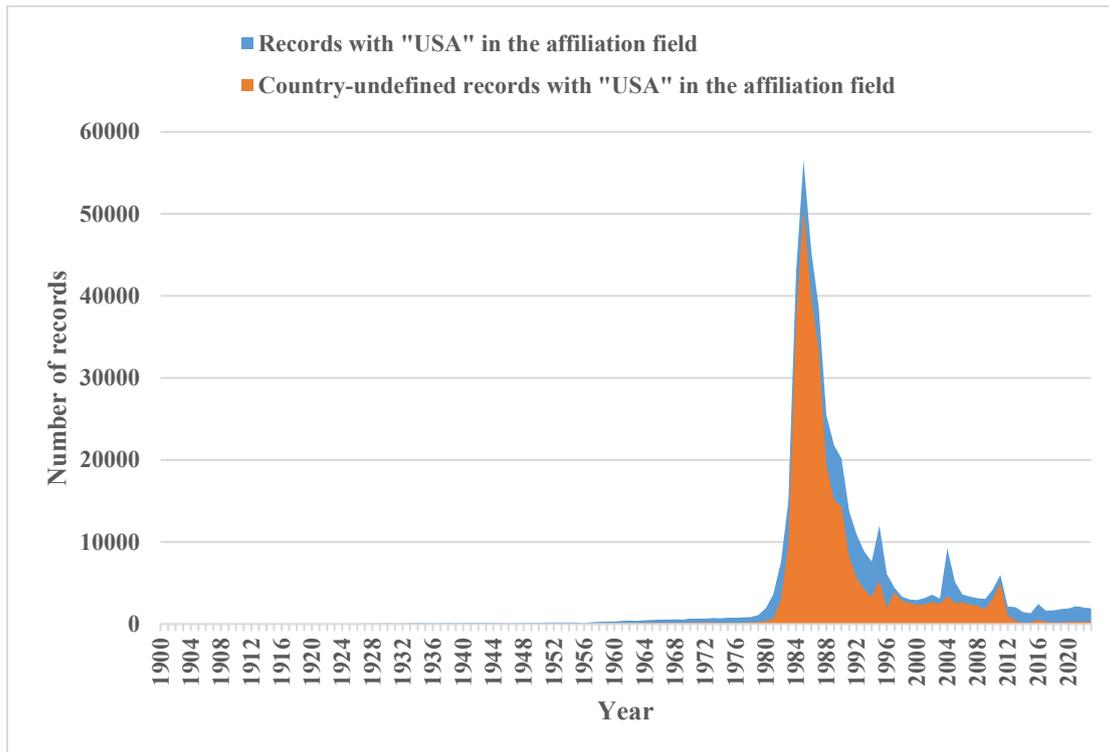

Figure 3 Dynamics of records with "USA" in the affiliation address field (all document types)

The third representative cause is the misspelled country/territory names in affiliation addresses. Some affiliated yet country-undefined records contain an affiliation country/territory portion in the affiliation fields in Scopus, however, the provided country/territory names are misspelled due to unknown reasons. For instance, a significant number of records contain the misspelled term "United Kingdon" in the affiliation field, suggesting an erroneous entry of "United Kingdom". The search query of "AFFIL("United Kingdon") AND PUBYEAR > 1899 AND PUBYEAR < 2024" retrieves over 9200 records in Scopus [9]. Among these publications, approximately 6400 are labeled as country-undefined by Scopus [10]. However, upon closer examination of these 6400 country-undefined records, we find that the majority of these records can be attributed to the United Kingdom due to Scopus' inability to recognize and correct misspellings in the country portion of these records' affiliation fields.

The fourth representative cause is that Scopus fails to correctly split and identify the clean affiliation addresses. The first scenario is the failure of Scopus to accurately split multiple pieces of affiliation addresses (for example, the record with doi: 10.1371/journal.pone.0074054) [11]. The composite affiliation address in this case hinders Scopus's capacity to accurately identify the affiliation country/territory information. The second scenario concerns the erroneous collection of non-affiliation address information by Scopus into the affiliation address field (such as email addresses, descriptions of authors and publications; for example, the record with doi:10.1017/S0140525X19001183). This leads to difficulties in the accurate identification of the affiliation country/territory information by Scopus due to the polluted or uncleaned affiliation address. For example, the search query "PUBYEAR > 1899 AND PUBYEAR < 2024 AND AFFIL ( China ) AND ( LIMIT-TO ( AFFILCOUNTRY , "Undefined" ) )" can hit about 30,000 country-undefined records with "China" in the affiliation field. By checking a sample of 100 country-undefined records in this group, we find about half of them belong to the above-mentioned two scenarios.

---

[9] Comparatively, less than 30 records can be retrieved for the same period by searching AD="United Kingdon" in the Web of Science Core Collection.
[10] About 2200 records are attributed to the country "United Kingdom" by Scopus. All these records contain "United Kingdom" in the country portion of the affiliation field.
[11] Scopus omits the third affiliation for this paper (Department of Aquatic Science and Assessment, Swedish University of Agricultural Science, Uppsala, Sweden).

## 5. Discussion

In this brief communication, we report a surprising finding that millions of affiliated but country-undefined records exist in Scopus. This huge number of "homeless" records is unacceptable for a widely used high-quality bibliographic database. Millions of "homeless" records are easily overlooked in Scopus-based literature retrieval, analysis, and evaluation. The primary and most evident cause is that a significant proportion of these publications' affiliation address fields are incomplete, specifically lacking the affiliation country/territory portion in Scopus. Besides, it is noteworthy that Scopus also classifies a considerable number of records with "nonstandard" or misspelled country/territory names as country-undefined. Another notable finding is that Scopus frequently fails to split and identify the clean affiliation addresses accurately, including failing to split multiple addresses accurately and incorporating non-address information into the affiliation address fields. While these records may contain "standard" affiliation country/territory names that are recognizable by Scopus, Scopus conservatively and even lazily labels a considerable number of them as country-undefined. All of these four factors, possibly with some undiscovered ones, add up to the millions of "homeless" records in Scopus[12].

The quality of metadata is the lifeblood of many established bibliographic databases (Besançon et al., 2024; Delgado-Quirós & Ortega, 2024; Krauskopf, 2017; Visser et al., 2021). Despite Scopus's self-designation as "a curated, high-quality bibliometric data source for academic research in quantitative science studies" (Baas et al., 2020), this study's finding of millions of affiliated but country-undefined records breaks this self-professed claim. Many of these records are highly cited papers (over 300 records have been cited over 2,000 times in Scopus) or published in world-leading journals (over 38,000 articles and reviews have been published by Nature and Science). Consequently, the absence of a designated response to this issue poses a significant threat to the reliability of Scopus-based literature retrieval analysis and evaluation. In contrast, the Web of Science Core Collection, Scopus's closest counterpart, has effectively addressed this issue. A mere 11 affiliated records have been identified as country-undefined in the Web of Science Core Collection over the past 124 years. A thorough examination of the metadata for these records in Web of Science reveals that eight of them have "*" in the country/territory fields, two have null values in the country/territory fields, and one has "Canada" at the end of the address field. The final special case is likely due to the absence of spaces after punctuation marks in the address field (doi:

---

[12] For example, some records with standard affiliation addresses collected in Scopus are also labeled as "country-undefined" due to unknown reasons (for example, the record with doi: 10.1109/TMM.2023.3340894).

10.2307/250920).

Scopus, a relative newcomer owned by the publishing giant Elsevier, has emerged as a contender to the established dominance of the Web of Science Core Collection (Singh et al., 2023; Zhu & Liu, 2020). Nevertheless, the quality of the metadata in Scopus falls short of the claimed "curated, high-quality" standard (Abalkina 2024; Franceschini et al., 2016; Krauskopf, 2019; Liu, 2020; Liu et al., 2024). While Scopus asserts a greater emphasis on precision over recall (Meester et al., 2016), the presence of millions of affiliated yet country-undefined records in Scopus is not consistent with the expected standards of a high-quality bibliographic database. It is acceptable to label records without affiliation addresses as country-undefined even for records without complete affiliation addresses. However, it is unacceptable to label many records with affiliation country/territory information collected in Scopus as country-undefined, especially for those with different variants of country/territory names.

## 6. Suggestions

As scholars, we identify a shocking phenomenon that millions of affiliated records in Scopus lack explicit country designations. Many of them are also with country/territory information collected in Scopus. We anticipate disseminating our findings to more users, data providers, and also owners of these bibliographic databases through this academic channel. To address this surprising and pressing issue, relevant stakeholders should take action to resettle these millions of "homeless" records in Scopus and reduce the potential impact.

As a responsible bibliographic database, Scopus must optimize its algorithm to effectively split multiple pieces of addresses and accurately identify the clean affiliation addresses. It is imperative for Scopus to be cognizant of the various variants of country/territory names, including those that may be misspelled with the aid of the provided affiliation addresses. Responsibly, Scopus should transform these variants and misspelled country/territory names into a standard and unique form. In a manner consistent with Web of Science, for records lacking country/territory information in the affiliation address field, Scopus should infer the affiliation country/territory information from the incomplete affiliation address field (such as the provided affiliation organization and geographical location information). The organization database of Scopus, as well as the knowledge graph Wikidata, are useful tools for addressing this issue (Nguyen et al., 2020). The inferred affiliation country/territory information should be added to the metadata of affected records in Scopus.

Secondly, academic publishers, including various journals, should request their authors to provide standard and complete author affiliation addresses. They should also provide accurate and complete metadata to bibliographic databases to enhance the reachability of their published works to users. Academic publishers can proactively address the metadata issues of their publications by collaborating with Scopus to resolve the existing problems. Collaborative effort is expected to enhance the reachability of their publications, ensuring that they are not overlooked in various analyses and evaluations, which are valued by authors.

Finally, users of Scopus must possess current knowledge of the database, including the most recent research results focusing on these bibliographic databases. It is imperative that users possess a comprehensive understanding of the database's limitations and deficiencies. Users of Scopus should fully consider the possible impact of these limitations and deficiencies, and develop corresponding countermeasures in literature retrieval, analysis, and evaluation. In the absence of effective countermeasures, disclosing the possible problems of the data to the readers is also a responsible option.

The deeper causes behind the phenomena found in this study may be more varied and difficult to pinpoint. Besides, the impact of millions of affiliated but country-undefined records is also difficult to measure. However, as long as stakeholders are informed and motivated to take action by this study, the aims of resettling millions of "homeless" records in Scopus and reducing the potential impact on Scopus-based literature retrieval, analysis, and evaluation can be achieved at a relatively low cost. It will greatly benefit the whole scientific community including authors and readers of JASIST.

**Conflict of interest statement**
The authors declare no conflicts of interest.